\documentclass[prl,twocolumn,showpacs,preprintnumbers,amsmath,amssymb]{revtex4}

\usepackage{graphicx}
\usepackage{dcolumn}
\usepackage{bm}

\begin{document}

\preprint{}

\title{Experimental determination of nucleation scaling law for small charged particles}

\author{Fabien Chirot}
\author{Pierre Labastie}
\author{S\'ebastien Zamith}
\author{Jean-Marc L'Hermite}
\affiliation{%
Laboratoire Collisions Agr\'egats R\'eactivit\'e, (UMR 5589 CNRS - Universit\'e Paul Sabatier Toulouse 3,\\
118 route de Narbonne, 31062 Toulouse Cedex 04 France
}%

\date{\today}

\begin{abstract}
We investigated the nucleation process at the molecular level.
Controlled sticking of individual atoms onto mass selected clusters
over a wide mass range has been carried out for the first time. We measured the absolute
unimolecular nucleation cross sections of cationic sodium clusters
$Na_n^+$ in the range $n$=25--200 at several collision energies. The
widely used hard sphere approximation clearly fails for small sizes:
Not only vapor-to-liquid nucleation theories should be modified but also, through
the microreversibility principle, cluster decay rate statistical models.
\end{abstract}
 \maketitle%
The formation andgrowth of droplets, clusters and, more generally,
very small particles is of considerable interest. Indeed these
phenomena are essential for instance in clouds formation or crystal
nucleation. The formation and growth of small particles can occur in
different ways. The formation can for example start from a seed
(heterogenous nucleation), in the presence of a buffer gas or simply
occur in a pure vapor (homogenous nucleation). The growth is, for
the smallest sizes, essentially governed by successive attachment of
single units whereas for bigger sizes coalescence dominates. A huge
amount of papers have been devoted to homogeneous nucleation, with
the aim of getting a correct quantitative theory of the phenomenon.
This aim has not been reached yet. A large majority of these
theories originates from the so-called "Classical Nucleation Theory"
(CNT), introduced by Becker and D\"{o}ring \cite{Becker1935}. It was
originally developed for neutral droplets but can be applied without
fundamental changes to ions (it is called in this case Classical
Ion-Induced Nucleation Theory, or CIINT) \cite{Kathmann2005,
Nadykto2004}. Ion-induced nucleation often predominates in nature
because ions have a higher sticking cross section (SCS) than neutral
particles. It has recently been demonstrated in the case of cirrus
clouds condensation \cite{Lee2003}. One crucial parameter in
nucleation is the growth rate, which depends on the size, defined as
the probability for a particle to grow by one unit per unit time. It
depends on the atomic flux and the SCS \cite{oxtoby1992,FordStat}.
For large enough particles, the SCS is given to an excellent
approximation by the hard sphere model. In the cluster size range
however, {\it i.e.} for very small particles made of tenth to 
hundredth units, this approximation is likely to fail. Furthermore 
all nucleation theories based on CNT define a so-called critical
size whose properties are crucial
\cite{FordStat,oxtoby1992,Oxtoby1994}. Unfortunately, in many
systems (water \cite{Wolk2001,Viisanen1993}, metals
\cite{bahadur2005}, organic compounds
\cite{Brus2006,Rudek1996,Strey1986}) the critical size lies in the
cluster size range where the hard sphere approximation may not be
relevant.

SCS's are also key parameters in the analysis of cluster decay rate.
As a matter of fact, in statistical models based on the
microreversibility principle (or detailed balance), the decay rate
is proportional to the SCS \cite{Weisskopf,Engelking1987}. Cluster
dissociation energies as deduced from their evaporation rates
\cite{diss_in_habbook} are thus influenced by sticking cross
sections.

The size dependence of the sticking cross-section has already been
addressed theoretically (see \textit{e.g.}
\cite{Vigue2000,Venkatesh1995}). However, while many experiments
have been devoted to the measurement of clusters evaporation rates
\cite{diss_in_habbook}, homogeneous SCS's of mass selected clusters
have never been measured to our knowledge. We present here the first
extensive measurement of clusters SCS's as a function of their size
and collision energy. The experimental setup has been described in
details elsewhere \cite{Chirot2006}. Sodium clusters are produced in
a gas aggregation source. They are ionized by a hollow cathode
discharge. They are then thermalized in a heat bath; their
temperature $T_i$ can be varied from 150 to 500~K. During cross
sections measurement, $T_i\approx 150$~K in order to avoid
evaporation at least until final products reach the detector. Wiley
Mc-Laren type electrodes operate a first mass selection. Next,
clusters enter an electrostatic device that reduces the kinetic
energy spread of the mass selected clusters. It allows to decrease
the kinetic energy to a few eV. Thermalized mass selected clusters
then propagate slowly across a cell containing a sodium vapor ($\sim
10^{-4}$~mbar for T=200~$^\circ$C). Finally, a second mass selection
determines the new size distribution at the output of the cell.

Our raw data are the relative number of clusters which undergo a
sticking $I/I_0$. This quantity is measured as a function of the
cluster size and kinetic energy. Assuming a Boltzmann velocity
distribution for the atoms, the SCS $\sigma_n$ is given by:
\begin{eqnarray}
\sigma_n &=& -\frac{\ln{I/I_0}}{l\rho}\left\{erf(\sqrt{a}) +
\frac{1}{2a}erf(\sqrt{a})+\frac{e^{-a}}{\sqrt{\pi a}}\right\}
\label{eq:one}
\end{eqnarray}
with $a=E_{k}/(nk_BT)$ and $l$ is the length of the cell, $\rho$ the
atomic density inside the cell, $n$ the size of the incoming
cluster, $T$ the temperature of the vapor and $erf$ is the error
function.  $E_k$ is the kinetic energy of the cluster in the
laboratory frame (in the following, kinetic energy refers to the
energy in the laboratory frame and collision energy to the energy in
the Center-of-mass frame). The density of atoms in the cell $\rho$
is deduced from the cell temperature through the vapor pressure
curve. We carefully measured the temperature at different positions
in the cell in order to determine as accurately as possible the
density of atoms from the lowest encountered temperature. The
density of atoms in the cell remains nevertheless the main source of
uncertainties. An error of 5~K on the cell temperature leads to a
relative error of 25\% on the determination of the cross section.
These uncertainties can shift the curves collectively, but do not
change the size dependence of the cross-section.

Note that under our experimental conditions evaporation is
completely negligible. We checked for evaporation effects on a few
sample masses: For Na$_{110}^+$ we varied $T_i$ from 150 to 250~K
and we did not see any evaporation effect. This is not surprising
since the lifetime for the product Na$_{111}^+$ is still 450~ms at
250~K. For Na$_{41}^+$ we varied $T_i$ from 150 to 200~K and found
no change in the SCS: the lifetime for Na$_{42}^+$ is still 63~ms at
200~K. For Na$_{55}^+$ we varied $T_i$ from 150 to 340~K and we did
see a change in the SCS after a rather long plateau. Evaporation
starts to affect the cross section at about 290~K. We find a
lifetime of 300~$\mu$s for the product Na$_{56}^+$ at this
temperature (Ec=20~eV) whereas the time of flight is about
130~$\mu$s. It is for sizes smaller than $n=25$ at $E_k$=20~eV and
$T_i$=150~K that the lifetime becomes comparable to the time of
flight. We can thus safely exclude the effect of evaporation for the
size range presented in our results.

Let us emphasize that contrary to previous experiments we work in
the single collision regime with very well defined initial
conditions (size, temperature and collision energy) and that we
explored a wide range of sizes ($n=25-200$). In the few studies
where reaction cross section of mass selected particles are measured
as a function of size or collision energy reactants underwent
evaporation and/or the size range was quite small
\cite{Orii,Ichihashi2005,Balteanu2003}.

\begin{figure}
\includegraphics[width=8cm]{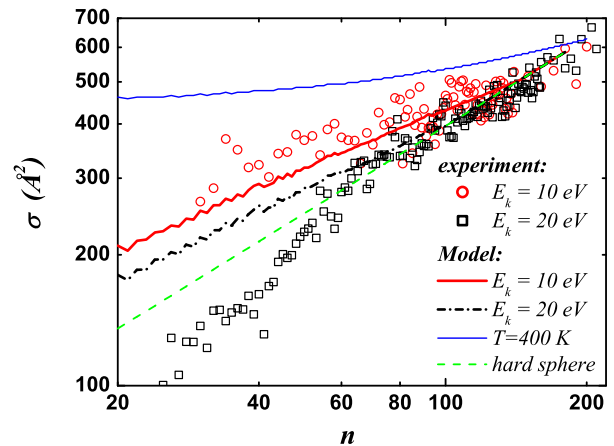}
\caption{\label{fig:1} Log-Log plot of the SCS as a function of
cluster size at two kinetic energies. Experimental results are
compared to a Langevin model and a hard sphere model. The calculated
mean cross sections in a vapor at 400 K are also shown.}
\end{figure}

\begin{figure}
\includegraphics[width=8cm]{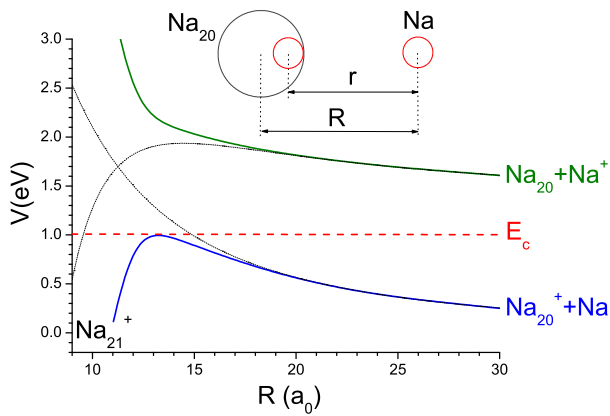}
\caption{\label{fig:2} Effective interaction potentials in the
harpooning model in the case of the of $Na_{20}^++Na\rightarrow
Na_{21}^+$ sticking reaction. Dotted lines: diabatic potentials.
Solid lines: adiabatic potentials. The coupling is taken as the
hopping integral in $Na_2^+$ at the distance $r$. The impact
parameter is equal here to its critical value $b_{max}$  (see
text).}
\end{figure}

SCS's have been measured at two kinetic energies. Figure \ref{fig:1}
presents our experimental results for $E_k$=10~eV and 20~eV. The
most noticeable feature is that the two curves are different for
small sizes and merge at about $n=80$. From this size on, they
follow the geometrical scaling law in $n^{2/3}$ predicted by the
hard sphere model (Figure \ref{fig:1}): $\sigma_{geo} = \pi
R^2n^{2/3}$, with $R=2.4$~{\AA} (the Wigner-Seitz radius for sodium
is 2.1~{\AA}). For smaller sizes, the cross section departs from the
hard sphere model. It drops below the geometric cross section for
$E_k=20$~eV whereas it increases for 10~eV. For sizes smaller than
80, the cross section depends on the collision kinetic energy: it
decreases for increasing kinetic energy. Figure \ref{fig:1} also
displays cross sections calculated in the frame of a Langevin model
\cite{Langevin,Orii}. The cluster-atom interaction is described by
charge-induced dipole potentials. Two potentials are involved,
corresponding respectively to $Na_n^+ + Na$ and $Na_n + Na^+$ (see
Fig. \ref{fig:2}). These two states are asymptotically separated by
the ionization potential difference $\Delta IP$ between atom and
cluster. In this frame, the sticking can be described as a
harpooning mechanism: due to the much larger polarisability of the
cluster compared to the atom, an electron transfer from the atom
onto the cluster is energetically favored. The adiabatic potentials
presented in Figure~\ref{fig:2} are calculated using as a coupling
the hopping integral in $Na_2^+$ \cite{Spi}.
From the Landau-Zener formula, the propagation is found to take
place on the ground state potential curve. In the Langevin model,
the largest reactive impact parameter $b_{max}$ is reached when the
maximum of the effective potential is equal to the collision energy.
The cross-section is then given by $\sigma_n = \pi b_{max}^2$. It is
taken as geometrical when the Langevin cross section becomes smaller
than the geometrical one. This simple model gives satisfactory
results for $E_k=10$~eV. However, at $E_k=20$~eV, calculated values
disagree with the experiment: the model is unable to account for
experimental cross sections being clearly below the geometrical
ones. More elaborated models exist
\cite{Marlow1982,Kummerlowe2005,Kasperovich2000,Hansen2004,Gluch2004}
but they would also give a cross-section larger than the geometrical
one. Moreover we have a charge-induced dipole interaction whereas
\cite{Marlow1982} deals with neutral particles and
\cite{Kummerlowe2005} consider molecules with permanent dipole.

One possible explanation for this behavior is the presence of non
reactive excited states of the cluster. Indeed since the experiment
is done at fixed kinetic energies in the laboratory frame, the
collision energy $E_c$ increases as the size $n$ decreases:
$E_{c}=\frac{E_{k}}{n+1} + \frac{3k_{B}T}{2}\frac{n}{n+1}$. Non
reactive excited states that can be reached as the collision energy
increases are likely to be responsible for the drop in the cross
sections for small $n$. Preliminary calculations based on a
two-center jellium model show the appearance of excited states
corresponding to the first excitations of $Na_n^+$. These excited
states lay between the two curves shown in Figure~\ref{fig:2} and
might be responsible for quenching towards non-reactive channels
\cite{Spi}. Our model thus applies only at low collision energy.
Practically however, nucleation and evaporation processes involve
thermal range collision energies which can be considered as small in
the sense defined here. Let us now examine in more details the size
dependance of $\sigma_n$. First note that there are no strong
structures. We introduced in our model the experimental values for
the $IP$ \cite{brechignac2000} and for the polarisabilities
\cite{Rayane,Kresin2001}. Although these values exhibit the well
known variations due to electronic shells closure \cite{Knight}, no
magic numbers clearly appear in the cross sections. This is
consistent with the unstructured mass spectra observed when clusters
are produced at low temperature \cite{baletto2005,Borggreen2000}.

The monotonic component does not follow a geometrical scaling law in
$n^{2/3}$. Actually, this behavior is not really surprising. It has
been experimentally observed in heterogeneous sticking reactions of
small mass selected clusters \cite{Ichihashi2005}. The effect on
homogeneous cluster growth of a discrepancy from the hard sphere
approximation had already been theoretically suggested in the case
of neutral particles interacting through a Van-Der-Waals interaction
\cite{Vigue2000}. In this case, $\sigma_n$ scales as $n^{1/3}$.
Surprisingly, this kind of effect had never been analyzed nor {\it a
fortiori} experimentally demonstrated in the case of charged
clusters.
This non geometrical sticking law has consequences for instance on
clusters decay rates and kinetic energy release distributions (KERD)
which are generally analyzed in the frame of statistical physics
models \cite{diss_in_habbook}. Most of the models used today are
based on the microreversibility (or detailed balance) principle,
whose basic assumption is summarized by the following relation
\cite{diss_in_habbook,Engelking1987}:
\begin{eqnarray}
\frac{\gamma_n(E_n,\varepsilon)}{\beta_{n-1}(\varepsilon)} = -\frac{\Omega_{n-1}(E_n-D_n-\varepsilon)\times\Omega_{atom}(\varepsilon)}{\Omega_n(E_n)}%
\label{eq:2}
\end{eqnarray}
where $\gamma_n$ is the cluster decay rate, $\beta_{n-1}$ the
sticking rate, $E_n$ the internal energy, $D_n$ the dissociation
energy and $\varepsilon$ the kinetic energy of the evaporated atom.
$\Omega_n$, $\Omega_{atom}$ and $\Omega_{n-1}$ are the densities of
state respectively of the parent cluster, the evaporated atom and
the product cluster. $\beta_{n-1}$ is related to the sticking cross
section by the relation $\beta_{n-1}=\sigma_{n-1}\rho v$ , where
$\rho$ is the volumic density of atoms and $v$ the atom velocity.
The point we lay the stress on here is that in this microreversible
approach the decay rate $\gamma_n$ or the KERD
\cite{Hansen2004,Gluch2004} are directly related to the sticking
cross section $\sigma_{n-1}$.

Dissociation energies are deduced from decay times $\tau_n$ at
cluster temperature $T$ through approximations of relation
\ref{eq:2} that can be written to a good accuracy as
\begin{eqnarray}
\tau_{n}^{-1} = \sigma_{n-1} A_n \exp{\left(-D_n/(k_BT)\right)}%
\label{eq:3}
\end{eqnarray}
$A_n$ is a prefactor independent of $\sigma_n$  whose exact
expression depends on the model. Dissociation energies are deduced
from $\tau_n$ by inverting relation \ref{eq:3}. For sodium clusters,
the experimental temperature must be of the order of 400K in order
to observe dissociation in the typical accessible time range
\cite{Brechignac1989,Borggreen2000}. At this temperature, the cross
section becomes geometrical at about {\it n$\approx$}200 as deduced
from our model (see fig.~\ref{fig:1}). One can estimate the order of
magnitude of the error made by using a geometrical sticking cross
section $\sigma_{geo}$. Replacing $\sigma_{geo}$ by our estimated
value at 400K (see fig.~\ref{fig:1}) leads to $\Delta D = \pm 4\%$.
This error is at least similar to experimental uncertainty and
should be corrected in future experiments on sodium. Unfortunately,
a general form of the correction cannot be inferred here since it is
strongly system-dependent, essentially through the energy shift and
the coupling between the two (or more) electronic states involved.

Nucleation in the frame of the CNT also depends in principle on the
SCS scaling law. Nucleation theories aim to estimate the nucleation
rate $J_i$ defined as the number of droplets of size $i$ created
from gas phase per unit time and volume. Although a number of
different versions have been developed since the work of Becker and
D\"{o}ring, the basic assumptions remains the same
\cite{FordStat,oxtoby1992}. CNT is derived from the population
equations:
\begin{eqnarray}
\frac{\partial n_i}{\partial t} =
\beta_{i-1}n_{i-1}-\gamma_in_i-\beta_in_i+\gamma_{i+1}n_{i+1} =
J_{i-1}-J_i \label{eq:5}
\end{eqnarray}
where $\beta$ and $\gamma$ are respectively growth and decay rates.
The CNT considers the steady state solution such as the flux $J$ is
constant for all $i$. This solution $J_M$ is valid in a medium time
range. Growth at very short delays will briefly be examined later.
For very large delay, the nucleation phenomenon is qualitatively
different, notably involving coalescence. Moreover, gas depletion
prevents particles from growing indefinitely. All CNT-based theories
arrive at an essentially common expression
\cite{FordStat,oxtoby1992,Shneidman2005}:
\begin{eqnarray}
J_M=K\beta_{i^*}\exp{\left( -\Delta G(i^*)/(k_BT) \right)}
\label{eq:6}
\end{eqnarray}
where $k_B$ is the Boltzmann constant, $T$ the temperature, $\Delta
G(i*)$ the formation free energy of the critical cluster of size
$i^*$. The critical size $i^*$ is defined by $\frac{\partial \Delta
G}{\partial i}|_{i^*}=0$, which is equivalent to
$\frac{\beta_{i^*}}{\gamma_{i^*+1}}=1$~\cite{FordStat}. The
prefactor $K$ does not depend on the cross section. $K$ includes the
so-called Zeldovich factor \cite{Zeldovitch} which does not depend
on the sticking scaling law provided that it is a smooth function of
the size \cite{oxtoby1992,FordStat}. Relation \ref{eq:2}, thus the
microreversibility principle, makes $i^*$ independent of the SCS
scaling law. This had already been pointed out by Vasil'ev and Reiss
\cite{Vasilev}. Finally, the nucleation rate in the constant flux
regime is modified only linearly through $\beta_{i^*}$. In CNT,
$\beta_{i^*}$ is always calculated under the assumption of hard
sphere collisions. In CIINT, however, many authors consider that the
stronger interaction between charged clusters and atoms leads to an
enhancement of $\beta_{i^*}$. They introduce a so-called Enhancement
Factor (EF)\cite{Nadykto2004,Vasilev}. If the vapor molecule has no
permanent dipole, EF is generally considered as negligible, as far
as clusters bear a single charge \cite{Fisenko}. However, when
evaluating EF, only the interaction between the charged cluster and
the neutral particle is considered. It is demonstrated here that, at
least in the case of sodium, the harpooning mechanism strongly
modify the SCS's. An example of EF(n) at 400~K is readily deduced
from fig.\ref{fig:1}: it is the ratio
$\sigma_{400~K}$/$\sigma_{geo}$. EF depends on the system, the
temperature, the coupling between the electronic states. So we
cannot derive any general rule for estimating EF. Nevertheless, we
have shown that more attention should be paid to this factor.

Let us focus now briefly on short times. There is a so-called lag
time $t_L$ that characterizes the time spent before the steady state
is reached. The nucleation rate $J$ follows approximately the law
$J(t)\approx J_M(1-\exp{(-B\beta_{i^*}t)})$, where the prefactor $B$
do not depend on the SCS \cite{Kelton1983}. Here the change could be
more significant; however, the effect on the steady-state nucleation
rates is not significant since the lag time by itself is very short
in vapor-to-liquid transformations \cite{Kelton1983}. Far from
equilibrium phenomena may therefore be more influenced by a change
in the sticking scaling law. Cluster nucleation in a supersonic
expansion is an example of such phenomena. As mentioned above,
Vigu\'e \textit{et al} already studied the case of neutral clusters
\cite{Vigue2000}. The final average cluster size $n_f$ can be
represented by empirical scaling laws of the type $n_f\approx
\Gamma^{\frac{1}{1-\alpha}}$ for a sticking law in $n^\alpha$. The
scaling parameter $\Gamma$ introduced by Hagena only depends on the
geometry, the pressure and the temperature in the source
\cite{Hagena1987}. The example calculated at thermal energy on
figure \ref{fig:1} shows that $\alpha$ can be close to 0. The change
in $n_f$ is obviously not negligible.

In summary, absolute SCS of mass selected sodium clusters have been
measured for the first time. The non geometrical sticking law that
depends on the collision energy is shown to influence not only
nucleation but also cluster evaporation measurement results.
\begin{acknowledgments}
We gratefully acknowledge F. Spiegelman for communicating
unpublished results. 
\end{acknowledgments}
\bibliography{controlledstickingetal}
\end{document}